\documentstyle[aps,preprint,tighten,eqsecnum]{revtex}

\input epsf

\newcommand \be {\begin{equation}} 
\newcommand \ee {\end{equation}}
\newcommand \ba {\begin{array}{c}} 
\newcommand \ea {\end{array}}
\newcommand \bea {\begin{eqnarray}} 
\newcommand \eea {\end{eqnarray}}

\def\oppropto{\mathop{\propto}} 
\def\opsimeq{\mathop{\simeq}}

\def\operarrow{\mathop{\longrightarrow}}
\def\opsimeq{\mathop{\simeq}} 

\begin{document} 
\draft 
\preprint{Saclay T97/161}

\title{\bf PHASES OF RANDOM ANTIFERROMAGNETIC SPIN-1 CHAINS}

\author{C. Monthus$^1$, O. Golinelli$^2$ and Th. Jolic\oe ur$^2$}

\address{$^1$Division de Physique Th\'eorique, IPN, Bat. 100,
91406 Orsay, France}

\address{$^2$Service de Physique Th\'eorique, CEA Saclay, 91191
Gif-sur-Yvette, France\\ e-mail : monthus@ipno.in2p3.fr, golinel,
thierry@spht.saclay.cea.fr} 

\date{today}

\maketitle 
\begin{abstract}
We formulate a real-space renormalization scheme that 
 allows the study of the effects of bond randomness
in the Heisenberg antiferromagnetic spin-1 chain.
There are four types of bonds that appear during the 
renormalization flow. We implement numerically the decimation 
procedure. We give a detailed study of the
probability distributions of all these bonds in the phases
that occur when the strength of the disorder is varied.
Approximate flow equations are obtained in the weak-disorder 
regime as well as in the strong disorder case where the 
physics is that of the random singlet phase.
\end{abstract}

\section{Introduction}

The effect of quenched impurities on the physics of one-dimensional
spin systems is an important and unsolved problem. Many spin chains
can be doped chemically and this creates some kind of disorder in the
system. In addition the spin-1/2 chain is equivalent to a system of
spinless fermions through the Jordan-Wigner transformation. This
means that the problem of interacting spinless fermions in a 
disordered
potential is equivalent to a random spin chain problem. There are 
not
so many techniques that allow the study of these systems. The
real-space renormalization group is prominent among them. Some time
ago a pioneering study by Ma and Dasgupta\cite{Ma79} showed that the
spin-1/2 Heisenberg antiferromagnetic chain with bond randomness is
in a so-called random-singlet phase. In this phase the spins are
locked into singlets that extend over arbitrarily long distances, in
a pattern dictated by the bond distribution. It has been realized
recently that the results of their renormalization procedure are in
fact {\it exact}\cite{Fis94}. This random-singlet phase may capture
the physics of higher-dimensional disordered systems\cite{BL82}.

In the spin-1/2 case, the random-singlet phase appears for various
kind of disorder and in a wide regions of the phase diagram when one
adds XXZ anisotropy. This results from the study of the weak disorder
regime by bosonizing the spin chain\cite{Doty}.

The spin-1 Heisenberg antiferromagnetic chain has a physics which is
vastly different in the pure case. There is a gap to spin excitations
and a finite spin-spin correlation length. These features can be best
understood by consideration of a hidden topological
order\cite{Rom87,Gir88}. In fact the ground state of the spin-1 chain
has a hidden long-range order that can be measured only by use of a
nonlocal correlation function, the so-called string order parameter.
It is a natural question to ask what happens to these peculiar
features under the influence of disorder. In fact the original
Ma-Dasgupta renormalization scheme requires a broad enough bond
distribution to work\cite{Boe96}. So  more complex schemes have been
proposed\cite{hymanPhD,HY,lettre}. As a function of the disorder
strength, it has been established that there is a phase transition
between a low-disorder gapless phase with hidden order and a strong
disorder phase which is the random-singlet phase of Ma and Dasgupta
(gapless and no hidden order).

In this paper, we give a detailed construction of a renormalization
scheme suited to the study of the spin-1 chain. We generalize the
Ma-Dasgupta decimation procedure by keeping more degrees of freedom.
A brief account has been given in Letter form\cite{lettre}. Here we
obtain explicit flow equations that are valid deep inside each of the
phases that appear. We are able to follow the spin populations as a
function of the renormalization scale as well as the evolution of
distribution functions of the various kinds of bonds that appear. In
section II, we define the renormalization scheme. In section III, we
study the weak-disorder phase of the spin-1 chain. Section IV
contains our results for the strong disorder regime. The critical
regime is studied in section V and section VI contains our
conclusions.

\section{Real Space Renormalization procedure for disordered
antiferromagnetic spin-1 chain}

In this section, we explain how to obtain a real space
renormalization scheme adequate to study the disordered
antiferromagnetic spin-1 chain.

\subsection{The Ma-Dasgupta real-space renormalization in the
spin-1/2 case} \label{Ma}

Ma and Dasgupta have introduced a real-space renormalization
procedure for the random antiferromagnetic spin-1/2 chain described
by the Hamiltonian 
\be 
H= \sum_i J_i \vec S_i . \vec S_{i+1} \ \ ,
\ee 
where $\{\vec S_i\}$ are quantum spin-1/2 operators and $\{J_i\}$
are positive random variables distributed with some probability
distribution $P_0(J)$. Suppose that $J_1$ is the largest coupling in
the chain. The one-bond Hamiltonian, 
\be 
h_0= J_1 \vec S_1 . \vec
S_2={J_1 \over 2} \left[\left( \vec S_1+ \vec S_{2} \right)^2 -\vec
S_1^2-\vec S_{2}^2 \right] ={J_1 \over 2} \left[\left( \vec S_1+ \vec
S_{2} \right)^2 -{3 \over 2} \right] \ \ , 
\ee 
admits two energy
levels labeled by $s=0,1$ 
\be 
e_{s}={J_1 \over 2} \left[s(s+1)- {3
\over 2} \right] 
\ee 
the level $e_{s}$ being $(2s+1)$ times
degenerate : $e_0=-{3 \over 4} J_1$ represents the singlet, and
$e_1={1 \over 4}J_1$ the triplet. At energies much lower than $J_1$,
the spins $\vec S_1$ and $\vec S_2$ will therefore be frozen into the
singlet state $s=0$. The decimation procedure consists in eliminating
the spins $\vec S_1$ and $\vec S_2$, and in replacing the four spin
segment Hamiltonian $H_{0,1,2,3}$ involving the decimated spins $\vec
S_1$ and $\vec S_2$ 
\be 
H_{0,1,2,3}= h_0+ h_1 \qquad \ 
\hbox{where} \
\ h_1= J_0 \vec S_0 . \vec S_1+ J_2 \vec S_2 . \vec S_3 
\ee 
by the
effective Hamiltonian for the remaining spins $\vec S_0$ and $\vec
S_3$ 
\be 
H_{0,3}^{eff}=E_{0,3}'+ J_0' \vec S_0 . \vec S_3 
\ee 
which
is meant to reproduce the four low-energy states of $H_{0,1,2,3}$
which are separated from the other twelve states of $H_{0,1,2,3}$ by
a big gap of order $J_1$. Using second order perturbation theory to
treat $h_1$ gives 
\be 
E_{0,3}'=-{3 \over 4} J_1 -{3 \over {16 J_1}}
(J_0^2+J_2^2) \ee and \be J_0'={{J_0 J_2} \over {2J_1}}
\label{marule} 
\ee 
The same procedure may be iterated and
successively applied to the new strongest bond of the chain. This
defines a flow for the probability distribution of couplings
$P(J,\Omega)$ where $\Omega$ is the current strongest coupling
\cite{Ma79} 
\be 
-{{\partial P(J,\Omega)} \over {\partial
\Omega}}=P(\Omega,\Omega) \int_0^{\Omega} dJ_a \int_0^{\Omega} dJ_b \
P(J_a,\Omega) \ P(J_b,\Omega) \ \delta\left(J-{{J_a J_b} \over
{2\Omega}} \right) 
\ee 
This flow equation has to be supplied by some
initial condition $P(J, \Omega_0)$. Fisher has shown \cite{Fis94}
that, for generic initial conditions, in the reduced variables
$\Gamma=\ln \left({{\Omega_0} \over \Omega} \right)$ and $z={1 \over
\Gamma} \ln \left({\Omega \over J} \right)$, the probability
distribution $R(z,\Gamma)$ of the variable $z$ flows towards the
unique fixed point $R^*(z)$ 
\be 
R(z,\Gamma) \operarrow_{\Gamma \to
\infty} R^*(z) \equiv \theta(z) \ e^{-z} 
\ee 
where $\theta$ is the
Heaviside step function. This so-called Random Singlet Fixed Point
corresponds to a power-law distribution in the original variables \be
P^*(J,\Omega)=\theta(\Omega-J) \ { {\alpha(\Omega)} \over {\Omega}} \
\left({J \over \Omega} \right)^{\alpha(\Omega)-1} \qquad \hbox{where}
\ \ \alpha(\Omega) \ \opsimeq_{\Omega \ll \Omega_0 } \ {1 \over {\ln
\left({ {\Omega_0} \over {\Omega}}\right)}} 
\ee 
for which two typical
bonds are typically much weaker than the strongest one $\Omega$. The
approximation involved in the use of perturbation theory to
obtain the rule (\ref{marule}) therefore becomes better and better as
the decimation proceeds, and the whole procedure is therefore
completely consistent even if the initial distribution is not broad.
The Ma-Dasgupta renormalization scheme is moreover very
appealing because it gives an interesting physical picture of the
random spin-1/2 chain : at low energy, the chain is made of pairs of
spins that are coupled together into singlets over arbitrarily long
distances, the long singlets bonds being typically much weaker than
the smaller ones.

\subsection{Renormalization of an AF bond between two spin-1}

The one-bond hamiltonian 
\be 
h_0= J_1 \vec S_1 . \vec S_2={J_1 \over
2} \left[\left( \vec S_1+ \vec S_{2} \right)^2 -\vec S_1^2-\vec
S_{2}^2 \right] ={J_1 \over 2} \left[\left( \vec S_1+ \vec S_{2}
\right)^2 - 4\right] 
\ee 
admits three energy levels labeled by
$s=0,1,2$ 
\be 
e_{s}={J_1 \over 2} \left[s(s+1)-4 \right] 
\ee 
the
level $e_{s}$ being $(2s+1)$ times degenerate : $e_0=-2J_1$
represents the singlet, $e_1=-J_1$ the triplet and $e_2=J_1$ the
quintuplet.

In the Ma-Dasgupta procedure, there are only two levels, and
``projecting onto the lowest level" is equivalent to ``projecting out
the highest level". Here these two possibilities are not equivalent.
The first possibility has already been considered in
refs.(\cite{hymanPhD},\cite{Boe96}) where it is shown that the
generalization of equation (\ref{marule}) describing the effective
coupling between $\vec S_0$ and $\vec S_3$ resulting from the
projection onto the singlet formed by $\vec S_1$ and $\vec S_2$,
reads 
\be 
J_0'={4 \over 3} {{J_0 J_2} \over {J_1}} 
\ee 
The
coefficient ${4 \over 3}$ being bigger than $1$, this rule is not
automatically consistent : indeed, the inequalities $J_0<J_1$ and
$J_2<J_1$ are not sufficient to imply that the new coupling $J_0'$ is
smaller than the decimated coupling $J_1$, in contrast with the rule
(\ref{marule}) concerning spin-1/2.  This procedure can however be
considered as qualitatively correct for very broad initial
randomness, where the cases which would produce a new coupling $J_0'$
bigger than the decimated coupling $J_1$ are statistically
negligible. So the strongly disordered antiferromagnetic spin-1
chains are described by the same random singlet fixed point already
found in the study of disordered spin-1/2 chains.

For weak initial randomness  however, this naive procedure cannot be
made consistent. We thus  generalize the Ma-Dasgupta procedure with
the interpretation of ``projecting out the highest level" instead of
``projecting onto the lowest level". More precisely for the
antiferromagnetic bond described by the Hamiltonian $h_0$, we project
out the quintuplet $e_2$ but to keep the singlet $e_0$ and the
triplet $e_1$ by replacing the two spin-1 $\vec S_1 $ and $\vec S_2$
by two spin-1/2 $\vec S_1' $ and $\vec S_2' $, and by replacing $h_0$
by the effective Hamiltonian 
\be 
h_0^{eff}= -{5 J_1 \over 4} +J_1
\vec S_1' . \vec S_2' \label{h0eff} 
\ee 
The four spin segment
Hamiltonian $H_{0,1,2,3}$ containing the old spins $\vec S_1 $ and
$\vec S_2 $ 
\be 
H_{0,1,2,3}= h_0+ h_1 \qquad \ \hbox{where} \ \ h_1=
J_0 \vec S_0 . \vec S_1+ J_2 \vec S_2 . \vec S_3 
\ee 
has to be
replaced by an effective Hamiltonian involving the spins $\vec S_0 $,
$\vec S_1' $, $\vec S_2' $ and $\vec S_3 $ 
\be H_{0,1,2,3}^{eff}=
h_0^{eff}+ h_1^{eff} .
\ee

If we use a first-order perturbation theory to treat $h_1$, we find
that the singlet of $h_0$ remains unchanged, whereas the degeneracy
of the triplet is lifted by the perturbation $h_1$. Using the
Wigner-Eckart theorem for vectorial operators, we find more
explicitly that the perturbation $h_1$ is equivalent to 
\be
h_1^{eq}=\left({1 \over 2} J_0 \vec S_0 +{1 \over 2} J_2 \vec S_3
\right) . (\vec S_1 +\vec S_2) 
\ee 
We therefore have to choose the
effective Hamiltonian 
\be 
h_1^{eff}= J_0 \vec S_0 . \vec S_1'+ J_2
\vec S_2' . \vec S_3 \label{h1eff} 
\ee 
since it is equivalent at
first-order perturbation theory, using again Wigner-Eckart theorem,
to the hamiltonian 
\be 
\left ({1 \over 2} J_0 \vec S_0 +{1 \over 2}
J_2 \vec S_3 \right) \left (\vec S_1' +\vec S_2' \right) 
\ee 
We have
now enlarged the initial space since the chain now contains not only
spin-1 but also spin-1/2. However it is possible to define a
decimation procedure that is ``closed" inside a particular set of
spin chains as we will see in the following.

\subsection{The real-space renormalization procedure }
\label{renorma}

We consider the enlarged set of spin chains described by the
Hamiltonian 
\be 
H= \sum_i J_i \vec S_i . \vec S_{i+1} 
\ee 
where the
spin $\vec S_i$ is a spin operator of size $s_i={1 \over 2}$ or
$s_i=1$, and where the couplings $\{J_i\}$ can be either positive or
negative, but have to satisfy the following constraint : for any pair
$\{i,j\}$ such that $i <j$, the classical magnetization of the
classical ground state of the segment $(i,j)$, must be smaller or
equal to one in absolute value 
\be 
\vert m_{i,j} \vert \leq 1
\label{magnecondition} 
\ee 
where the quantity $m_{i,j}$ reads 
\be
m_{i,j}= s_i + \sum_{n=i+1}^{j} s_n \times
sign\left[\prod_{p=i}^{n-1} (- J_p) \right] .
\ee

This condition for $j=i+1$ gives immediately that there are exactly
four types of bonds

\begin{itemize}

\item[1)] Link of type 1 : Ferromagnetic bond between two spin-1/2

\item[2)] Link of type 2 : Antiferromagnetic bond between two
spin-1/2

\item[3)] Link of type 3 : Antiferromagnetic bond between one spin-1
and one spin-1/2

\item[4)] Link of type 4 : Antiferromagnetic bond between two spin-1

\end {itemize}

Our decimation procedure is the following :

To each bond $\left( \vec S_i, \vec S_{i+1} , J_i \right)$ we
associate the energy difference between the higher state and the
lower state of the reduced Hamiltonian $J_i \vec S_i . \vec S_{i+1}$
\bea 
&\Delta_i= - J_i \qquad &\hbox{if the bond $i$ is of type 1} \\
&\Delta_i= J_i \qquad &\hbox{if the bond $i$ is of type 2} \\
&\Delta_i= {3 \over 2} J_i \qquad &\hbox{if the bond $i$ is of type
3} \\ &\Delta_i= 3 J_i \qquad &\hbox{if the bond $i$ is of type 4}
\label{gapdelta} 
\eea

We pick up the bond $\left( \vec S_{i_1}, \vec S_{i_2} , J_{i_1}
\right)$ corresponding to the strongest $\Delta_i$ of the chain. To
define the renormalization rule for this bond, we again divide the
four-spin Hamiltonian into 
\be 
H_{i_0,i_1,i_2,i_3}= h_0+ h_1 \qquad
\hbox{where} \ \ h_0=J_{i_1} \vec S_{i_1} \vec S_{i_2} \ \ \hbox{and}
\ \ h_1= J_{i_0} \vec S_{i_0} . \vec S_{i_1}+ J_{i_2} \vec S_{i_2} .
\vec S_{i_3} 
\ee 
and treat $h_1$ as a perturbation of $h_0$ to find
the effective Hamiltonian replacing $H_{i_0,i_1,i_2,i_3}$ when the
highest energy state of $h_0$ is removed. We have now to distinguish
the four types of bonds

\begin{itemize}

\item[Rule 1)] {\bf F bond between two spin-1/2 }

The hamiltonian $h_0=J_{i_1} \vec S_{i_1} \vec S_{i_2}$ admits two
energy levels : the triplet $e_1=-{{\vert J_{i_1} \vert} \over 4}$
and the singlet $e_0={{3 \vert J_{i_1} \vert} \over 4}$. The
perturbation $h_1$ lifts the degeneracy of the triplet, and using
Wigner-Eckart theorem, we find that $h_1$ is equivalent at first
order of perturbation theory to 
\be 
h_1^{eq}=\left({1 \over 2}
J_{i_0} \vec S_{i_0} +{1 \over 2} J_{i_2} \vec S_{i_3} \right) .
(\vec S_{i_1} +\vec S_{i_2}) 
\ee 
To eliminate the singlet state and
only keep the triplet state of $h_0$, we remove the two spin-1/2
$\vec S_{i_1}$ and $\vec S_{i_2}$ and replace them by a single spin-1
$\vec S_{i_1}^\prime$, and we replace $H_{i_0,i_1,i_2,i_3}$ by 
\be
H_{i_0,i_1',i_3}^{eff} =-{{\vert J_{i_1} \vert} \over 4} +{1 \over 
2}
J_{i_0} \vec S_{i_0} \vec S_{i_1}' +{1 \over 2} J_{i_2} 
\vec S_{i_1}' \vec S_{i_3} 
\ee

\item[Rule 2)] { \bf AF bond between two spin-1/2}

Here, we directly apply the Ma-Dasgupta procedure discussed in
\ref{Ma}: we remove the two spin-1/2 $\vec S_{i_1}$ and $\vec
S_{i_2}$ and replace $H_{i_0,i_1,i_2,i_3}$ by 
\be
H_{i_0,i_3}^{eff}=-{3 \over 4} J_{i_1} -{3 \over {16 J_{i_1}}}
(J_{i_0}^2+J_{i_2}^2)+ {{J_{i_0} J_{i_2}} \over {2J_{i_1}}} \vec
S_{i_0} . \vec S_{i_3} 
\ee

\item[Rule 3)] {\bf  AF bond between one spin-1 and one spin-1/2}

\label{Rule 3}

Suppose that $s_{i_1}=1$ and $s_{i_2}={1 \over 2}$. The hamiltonian $
h_0=J_{i_1} \vec S_{i_1} \vec S_{i_2}$ admits two energy-levels : the
doublet $e_{1/2}=-J_{i_1}$ and the quadruplet $e_{3/2}={J_{i_1} \over
2}$. At first order perturbation theory, Wigner-Eckart theorem gives
that, within the subspace of the doublet $s={1 \over 2}$, the
perturbation $h_1$ is equivalent to 
\be 
h_1^{eq}=\left(\alpha_1
J_{i_0} \vec S_{i_0} +\alpha_2 J_{i_2} \vec S_{i_3} \right) . (\vec
S_{i_1} +\vec S_{i_2}) 
\ee 
where the constants $\alpha_1$ and
$\alpha_2$ read 
\be 
\alpha_1={1 \over 2}
\left[1+{{s_{i_1}(s_{i_1}+1)-s_{i_2}(s_{i_2}+1)} \over {s(s+1)}}
\right]={4 \over 3} \quad \hbox{and} \ \ \alpha_2=1-\alpha_1=-{1
\over 3} 
\ee 
The renormalization rule is therefore the following : we
eliminate the spins $\vec S_{i_1}$ and $\vec S_{i_2}$, and replace
them by a single spin-1/2 $\vec S_{i_1}'$, and we replace
$H_{i_0,i_1,i_2,i_3}$ by the effective Hamiltonian 
\be
H_{i_0,i_1',i_3}^{eff} =-{J_{i_1} }+{4 \over 3} J_{i_0} \vec S_{i_0}
\vec S_{i_1} ' - {1 \over 3} J_{i_2} \vec S_{i_1} ' \vec S_{i_3} .
\ee

\item[Rule 4)] { \bf  AF bond between two spin-1}

In this case we apply the rule explained at the beginning of this
section (see eqs (\ref{h0eff})-(\ref{h1eff})) : we replace the two
spin-1 $\vec S_1 $ and $\vec S_2$ by two spin-1/2 $\vec S_{i_1}' $
and $\vec S_{i_2}' $, and we replace $H_{0,1,2,3}$ by an effective
Hamiltonian 
\be 
H_{i_0,i_1,i_2,i_3}^{eff}= -{5 J_{i_1} \over 4} +
J_{i_0} \vec S_{i_0} . \vec S_{i_1}'+ J_{i_1} \vec S_{i_1}' . \vec
S_{i_2}' +J_{i_2} \vec S_{i_2} ' . \vec S_{i_3} .
\ee

\end {itemize}


This renormalization procedure is entirely consistent from the point
of view of the progressive elimination of the highest energy degrees
of freedom : it is easy to show that in the four cases of
renormalization of a bond described above, all the energy scales
$\Delta_i$ of the new bonds are always smaller than the energy scale
$\Delta_{i1}$ of the bond that we renormalize.

It is also easy to check that this renormalization procedure is
``closed" inside the set of spin chains defined by the condition
(\ref{magnecondition}) : if we apply this procedure to an initial
chain belonging to this space, such as the random antiferromagnetic
spin-1 chain we are interested in, the effective chain always belongs
to this set of spin chains. In particular, spins higher than $1$
cannot appear through this renormalization scheme.

However, since this renormalization procedure is not purely based on
complete decimation of bonds, it introduces correlations between
bonds, so that it is impossible to write exact closed flow equations
for the probability distributions of couplings, in contrast with the
Ma-Dasgupta procedure. To study the properties of this
renormalization scheme, we have therefore performed numerical
simulations on spin-1 chains containing $N$ sites with periodic
boundary conditions ($N=2^{22}$ for example), whose initial couplings
$J_i$ are distributed according to probability distributions of the
following form 
\be 
P_d(J) ={1 \over d} \ \ \hbox{for} \ 1 \leq J \leq
1+d \qquad , \ \ \hbox{and $P_d(J)=0$ elsewhere} \label{PdJ} 
\ee 
The
parameter $d$ represents the strength of the initial disorder of the
couplings. For a given number of sites $N$, and a given initial
strength $d$ of the disorder, we have numerically implemented the
renormalization rules on a given number (typically 100) of initial
independent samples, to compute averaged quantities over these
different realizations of the initial disorder.
 It is convenient to use the variable :
 \be 
 \Gamma=\ln {{\Omega_0}
\over {\Omega}} ,
\ee 
where $\Omega$ is the current strongest $\Delta$
(see eq. \ref{gapdelta}) and $\Omega_0$ the initial strongest
$\Delta$. We have studied the flow of the following quantities : the
number $N(\Gamma)$ of effective spins $S=1/2$ and $S=1$ still 
present
at scale $\Gamma$ ; the proportion $\{ N_{(S=1) (\Gamma)}/ N
(\Gamma)\}$ of spins $S=1$ among the effective spins at scale
$\Gamma$ ; the proportions $\rho_i(\Gamma)=\{N_i(\Gamma)/N(\Gamma)\}$
of bonds of type $i=1,2,3,4$ at scale $\Gamma$; the probability
distributions $P_i(J,\Omega)$ of the coupling $J$ at scale $\Omega$
for the four types of bonds $i=1,2,3,4$. It is in fact more
convenient to study the probability distributions ${\cal P}_i (x,
\Gamma)$ of the reduced variable 
\be 
x=\ln \left({\Omega
\over{\Delta(J)}} \right) ,
\ee 
where $\Delta(J)$ is defined as in
(\ref{gapdelta}) 
\bea 
&\Delta(J)= - J \qquad &\hbox{for bonds of type
1} 
\\ 
&\Delta(J)= J \qquad &\hbox{for bonds of type 2} 
\\ 
&\Delta(J)=
{3 \over 2} J \qquad &\hbox{for bonds of type 3} 
\\ 
&\Delta(J)= 3 J
\qquad &\hbox{for bonds of type 4} 
\label{defxtype} 
\eea 
so that
the random variable $x$ varies in $(0,\infty)$ for any type of bonds.

\section{The weak disorder phase }

\subsection{Numerical results}

\begin{figure}[ht] 
\vglue 5.0cm
\centerline{\epsfbox{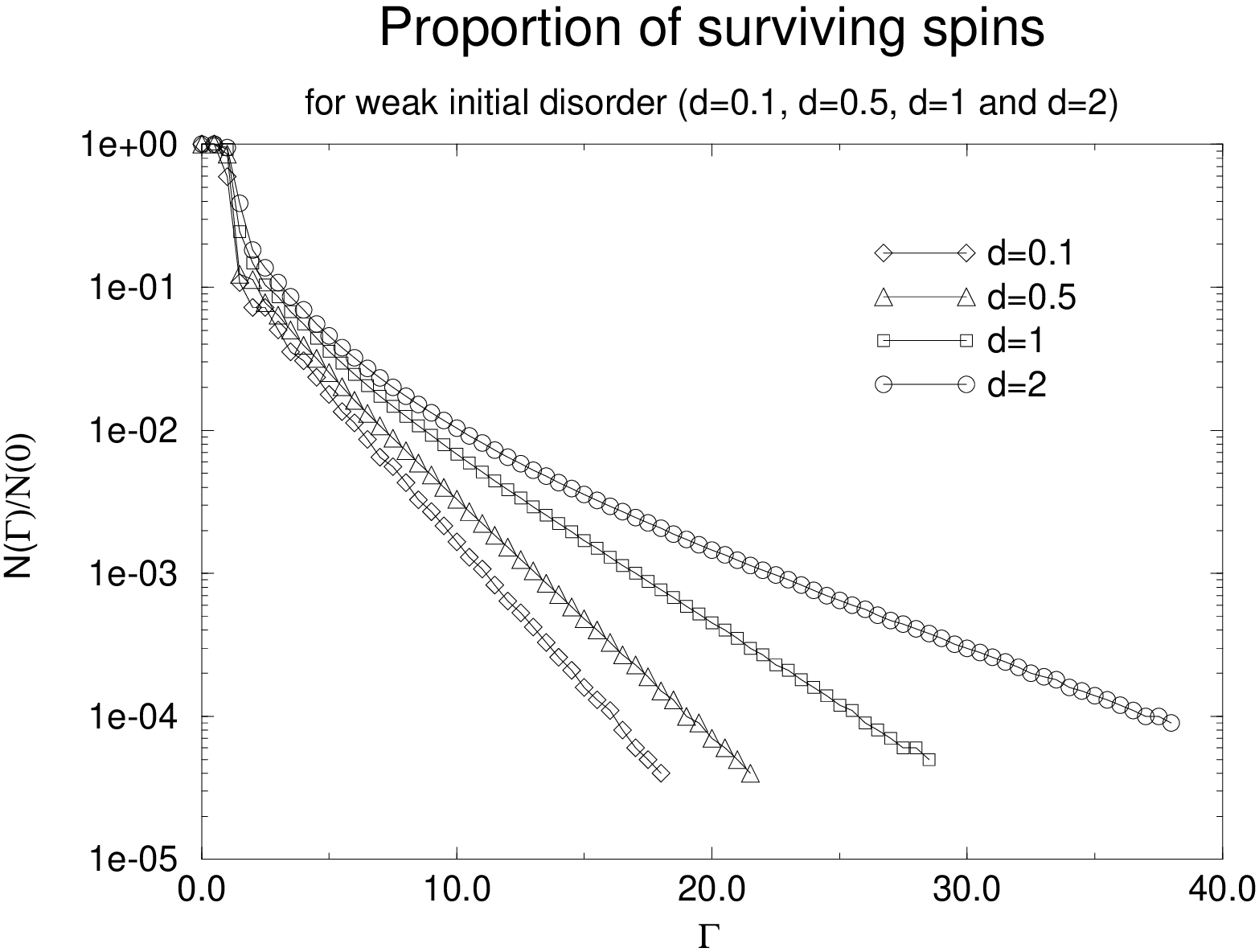}}
\vskip 2.0cm
\caption{
\label{ngaweak}
{Linear-Log plot of the proportion
${{N(\Gamma)} \over {N(0)}}$ of effective spins at scale $\Gamma$,
for weak initial disorder $d=0.1, d=0.5, d=1$ and $d=2$ : this
proportion decays exponentially (\ref{ngafaible}).}} 
\end{figure}

\begin{figure}[ht] 
\null
\vglue 5.0cm
\centerline{\epsfbox{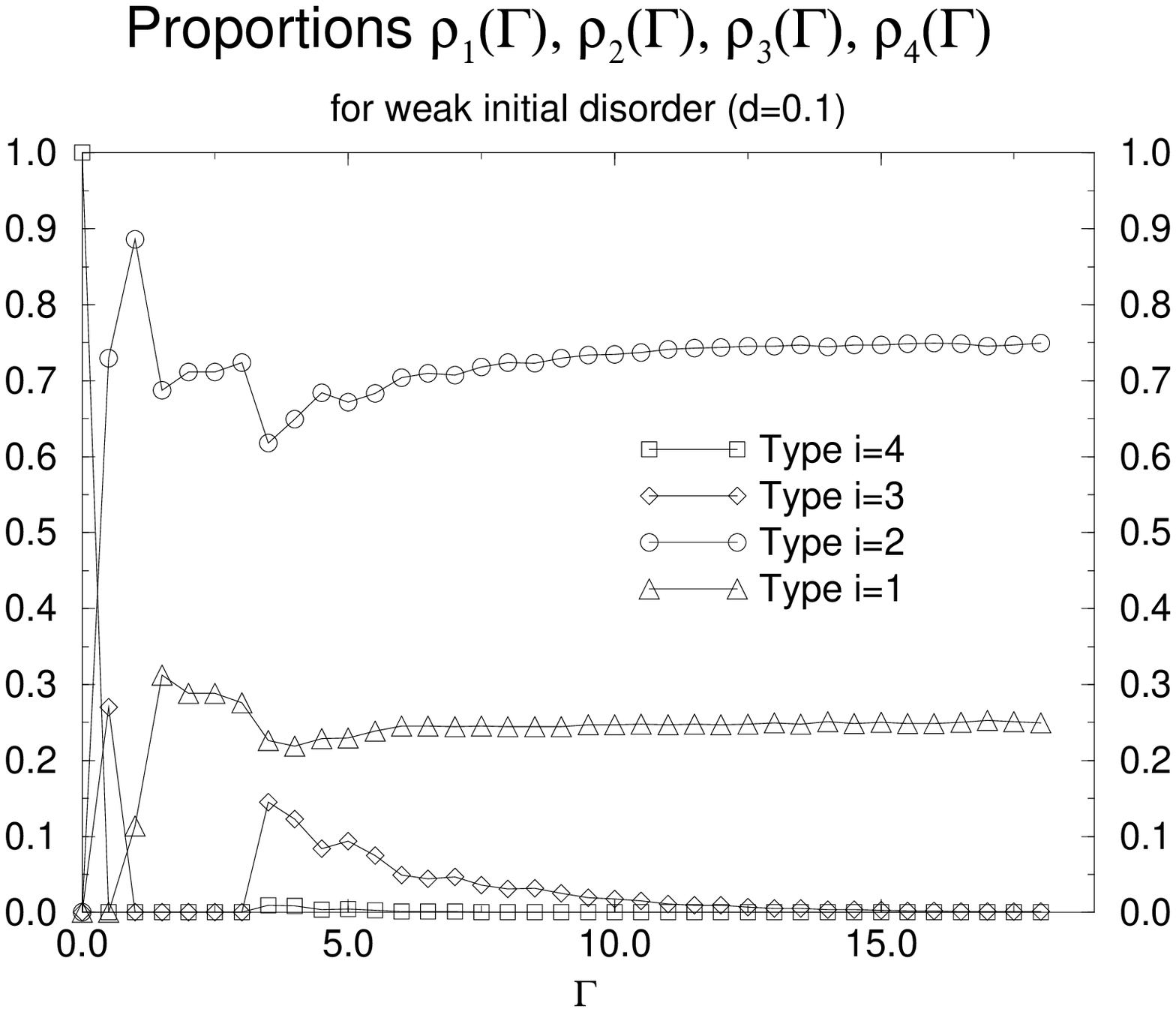}}
\vskip 2.0cm
\caption{
\label{rho01}
{ The proportions $\rho_i(\Gamma)$ of the four
types $i=1,2,3,4$ of bonds at scale $\Gamma$, for weak initial
disorder $d=0.1$ : they reach the asymptotic regime
(\ref{rhofaible}).}} 
\end{figure}

In the weak disorder phase, we find that the number $N(\Gamma)$ of
effective spins decays exponentially (see Fig \ref{ngaweak}) 
\be
N(\Gamma) \oppropto_{\Gamma \to \infty} e^{-\alpha(d) \Gamma}
\label{ngafaible} 
\ee 
where $\alpha(d)$ is a decreasing function of
the disorder $d$ that vanishes in the limit $d \to d_c^-$. 
As a consequence the magnetic susceptibility at temperature T
can be computed by summing Curie laws for the free spins at scale 
$\Omega = T$.
So we have~:
\be
\chi \oppropto {1\over T^{1-\alpha(d)}} .
\ee
The
proportions $\rho_i(\Gamma)$ of the four types of bonds reach a
stationary regime characterized by (see Fig \ref{rho01}) 
\be
\rho_1(\Gamma)\simeq 0.25 \qquad \ \rho_2(\Gamma) \simeq 0.75 \qquad
\ \rho_3(\Gamma) \simeq 0 \qquad \ \rho_4(\Gamma) \simeq 0
\label{rhofaible} 
\ee 
There are asymptotically only bonds of type 1
and bonds of type 2. This means in particular that there are only
effective spin-1/2 in the chain, and no more spin-1. Since two bonds
of type 1 cannot be neighbors according to the constraint
(\ref{magnecondition}), the even bonds and the odds bonds are not
equivalent, as in the effective model of Hyman and Yang \cite{HY} :
the ``even" bonds are all antiferromagnetic, whereas the ``odd" bonds
are either ferromagnetic or antiferromagnetic with equal probability.

\begin{figure}[ht] 
\null
\vglue 5.0cm
\centerline{\epsfbox{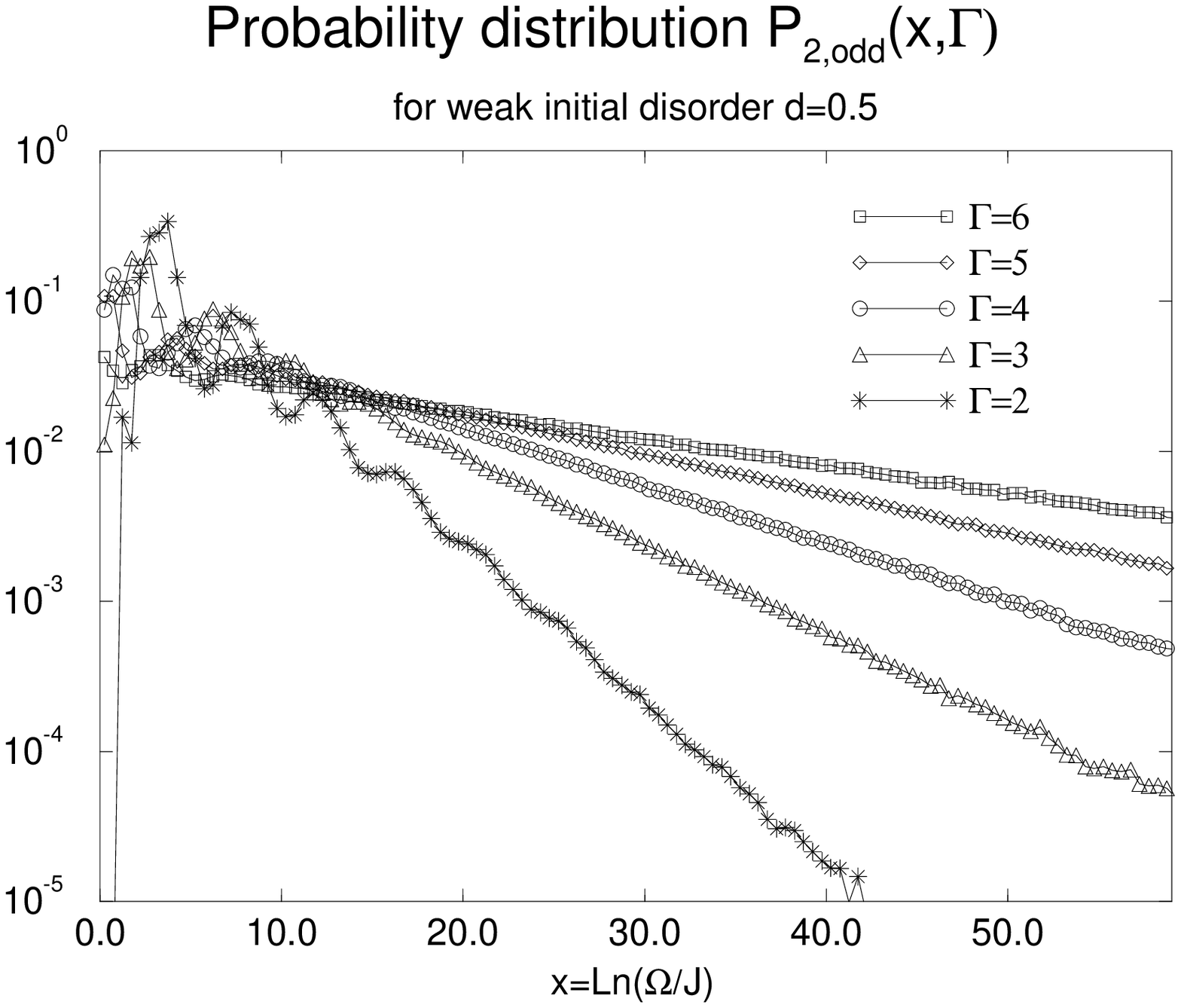}}
\vskip 2.0cm
\caption{
\label{p2oddd05}
{Linear-Log plot of the probability
distribution ${\cal P}_2^{odd} (x, \Gamma)$ for $\Gamma=2,3,4,5,6$,
for weak initial disorder $d=0.5$ : ${\cal P}_2^{odd} (x, \Gamma)$ is well described by the exponential form (\ref{poddnume}) with a
parameter $\alpha_o(\Gamma)$ that is found to decay exponentially
with $\Gamma$ (\ref{alphaodd}).}}
\end{figure}

It is necessary to introduce the probability distribution ${\cal
P}_2^{even} (x, \Gamma)$ for the couplings of the even bonds of type
2, and the probability distribution ${\cal P}_2^{odd} (x, \Gamma)$
for the couplings of the odd bonds of type 2. We find that ${\cal
P}_2^{even} (x, \Gamma)$ becomes stationary for large enough
$\Gamma$, and takes the form of an exponential distribution 
\be 
{\cal
P}_2^{even} (x, \Gamma) \simeq \alpha_e \ e^{- \alpha_e x}
\label{p2evennume} 
\ee 
where $\alpha_e$ is independent of $\Gamma$,
but depends on the value $d$ of the disorder, and is numerically very
close to the parameter $\alpha(d)$ characterizing the decay of
$N(\Gamma)$ (\ref{ngafaible}). The probability distributions ${\cal
P}_1 (x, \Gamma)$ and ${\cal P}_2^{odd} (x, \Gamma)$ coincide (up to
statistical fluctuations) and take the form of an exponential
distribution (see Fig. \ref{p2oddd05}) 
\be 
{\cal P}_1 (x, \Gamma)
\simeq {\cal P}_2^{odd} (x, \Gamma) \simeq \alpha_o(\Gamma)\ e^{-
\alpha_o(\Gamma) x} \label{poddnume} 
\ee 
where the parameter
$\alpha_o(\Gamma)$ decays exponentially \be \alpha_o(\Gamma)
\oppropto e^{- \alpha_e \Gamma} \label{alphaodd} 
\ee

As a consequence, for large enough $\Gamma$, the bond of the chain 
of
highest $\Delta$ (corresponding to smallest $x$) that is chosen to 
be
renormalized, is always an even bond of type 2. In the
renormalization operation (2), this even bond disappear together 
with
its two odd neighbors, and a new weak odd bond is produced. This
explains why the distribution ${\cal P}_2^{even} (x, \Gamma)$ for
even bonds remains stationary, whereas the distribution of couplings
of odd bonds becomes broader and broader in the variable $x$. This
weak disorder phase is therefore the same as the ``Haldane phase''
found by Hyman and Yang in their effective model introduced in
\cite{HY}, and is very similar to the random dimer phase found in 
the
study of random dimerized antiferromagnetic spin-1/2 chains
\cite{Hy96} : in the asymptotic regime, the chain is made of a set
of nearly uncoupled dimers.

\subsection{Approximate flow equations}

Assuming that the ``even" bonds are all of type 2, that the ``odd"
bonds are either of type 1 or of type 2 with equal probability, and
that the unique important process is the decimation of an even bond
according to the rule (2)

\bigskip \bigskip \hbox to 450pt {\hfill \vbox{\hsize=10pt
\centerline{$s_1={1\over 2}$}\par \centerline{$\bullet$}} \kern -7pt
\raise 2pt \vtop{\hsize=45pt \centerline{\hrulefill} \par
\centerline{$J_1$}} \kern -7pt \vbox{\hsize=10pt
\centerline{$s_2={1\over 2}$}\par \centerline{$\bullet$}} \kern -7pt
\raise 2pt \vtop{\hsize=45pt \centerline{\hrulefill} \par
\centerline{$J_2=\Omega$}} \kern -7pt \vbox{\hsize=10pt
\centerline{$s_3={1\over 2}$}\par \centerline{$\bullet$}} \kern -7pt
\raise 2pt \vtop{\hsize=45pt \centerline{\hrulefill} \par
\centerline{$J_3$}} \kern -7pt \vbox{\hsize=10pt
\centerline{$s_4={1\over 2}$}\par \centerline{$\bullet$}}
\hfill$\longrightarrow$\hfill \vbox{\hsize=10pt
\centerline{$s_1={1\over 2}$}\par \centerline{$\bullet$}} \kern -7pt
\raise 2pt \vtop{\hsize=135pt \centerline{\hrulefill} \par
\centerline{$J_1'={{J_1 \ J_3} \over {2\, \Omega}}$}} \kern -7pt
\vbox{\hsize=10pt \centerline{$s_4={1\over 2}$}\par
\centerline{$\bullet$}} \hfill } \bigskip \bigskip

it is possible to write approximate flow equations for the
probability distributions of the couplings are normalized according
to \be 1= \int_0^{\Omega} dJ \ P_2^{even}(J,\Omega) =\int_0^{\Omega}
dJ \ P_2^{odd}(J,\Omega) = \int_{-\Omega}^0 dJ P_1(J,\Omega) \ee It
is convenient to introduce the normalized distribution of all odd
bonds \be P^{odd}(J,\Omega) = {1 \over 2} \bigg(
P_2^{odd}(J,\Omega)+P_1(J,\Omega) \bigg) \qquad \hbox{for} -\Omega <J
<\Omega \ee The approximate flow equations for the probability
distributions $P_2^{even}(J,\Omega)$ and $P^{odd}(J,\Omega)$ then
read \be -{{ \partial P_2^{even}(J,\Omega) } \over {\partial \Omega
}} =P_2^{even}(\Omega,\Omega) \ P_2^{even}(J,\Omega) \label{floweven}
\ee \bea -{{ \partial P^{odd}(J,\Omega) } \over {\partial \Omega }} =
- P_2^{even}(\Omega,\Omega) P^{odd}(J,\Omega) \\ +
P_2^{even}(\Omega,\Omega) \ \int_{-\Omega}^{\Omega} dJ_1 \
P^{odd}(J_1,\Omega) \ \int_{-\Omega}^{\Omega} dJ_3 \
P^{odd}(J_3,\Omega) \ \delta \left( J-{{J_1 J_3} \over {2\,
\Omega}}\right) \label{flowodd} \eea

In the new variables $\Gamma=\ln {{\Omega_0} \over {\Omega}} $ and
$x=\ln \left({\Omega \over{ J }} \right) \in [0,+\infty)$, the flow
equation for ${\cal P}_2^{even}(x)$ admits stationary solutions of
exponential form \be {\cal P}_2^{even}(x) =\alpha_e \ e^{- \alpha_e
x} \label{p2even} \ee with undetermined constant $\alpha_e$, in
agreement with our numerical result (\ref{p2evennume}). With the last
change of variables 
\be 
x \longrightarrow z=\alpha_o(\Gamma) \ln
\left({\Omega \over{ \vert J \vert }} \right) 
\ee 
the flow equation
for the corresponding probability distributions $\tilde P^{odd}_1(z,
\Gamma)$ and $\tilde P^{odd}_2(z, \Gamma)$ admit the same stationary
solution 
\be 
\tilde P^{odd}_1(z, \Gamma) \simeq \tilde P^{odd}_2(z,
\Gamma) \operarrow_{ \Gamma \to \infty} e^{-z} \qquad \hbox{with} \
\alpha_o(\Gamma) \oppropto_{\Gamma \to \infty} e^{- \alpha_e \Gamma }
\ee 
where $\alpha_e$ is the number characterizing ${\cal
P}_2^{even}(x)$ (\ref{p2even}). We may also write the flow equation
for the total number $N(\Omega)$ of spins still present at scale
$\Omega$ 
\be 
-{{dN} \over {d\Omega}} = - P_2^{even}(\Omega,\Omega) \
N(\Omega) ,
\ee 
so that we obtain the following asymptotic behavior in
the variable $\Gamma$ :
\be 
N(\Gamma) \oppropto_{\Gamma \to \infty}
e^{- \alpha_e \Gamma } .
\ee

\section{The strong disorder phase }
\subsection{Numerical results}

\begin{figure}[ht] 
\null
\vglue 5.0cm
\centerline{\epsfbox{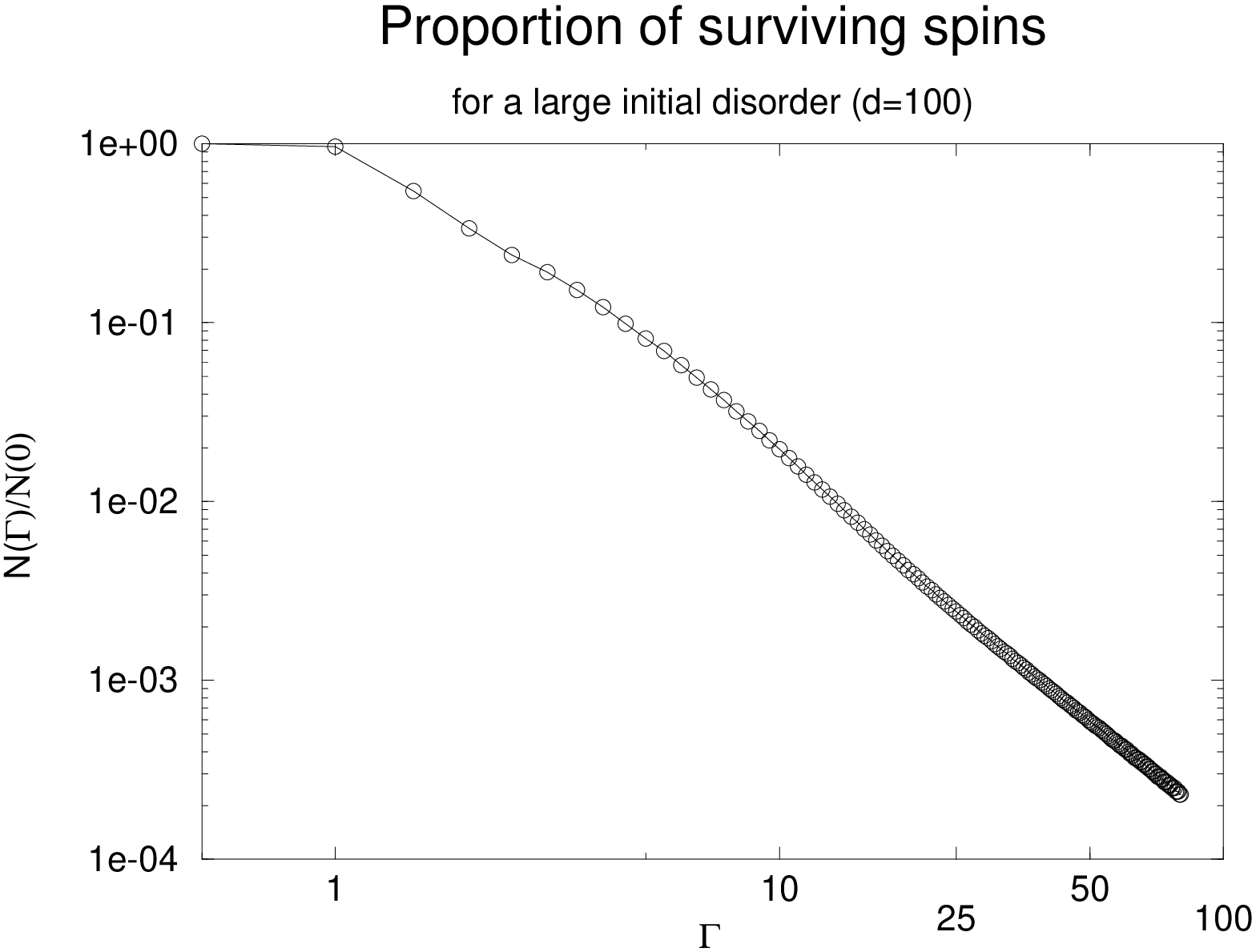}}
\vskip 2.0cm
\caption{
\label{nga100}
{Log-Log plot of the proportion ${{N(\Gamma)}
\over {N(0)}}$ of effective spins at scale $\Gamma$ for strong
initial disorder $d=100$ : this proportion follows the power-law
asymptotic behavior (\ref{ngafort}).}} 
\end{figure}

\begin{figure}[ht] 
\null
\vglue 5.0cm
\centerline{\epsfbox{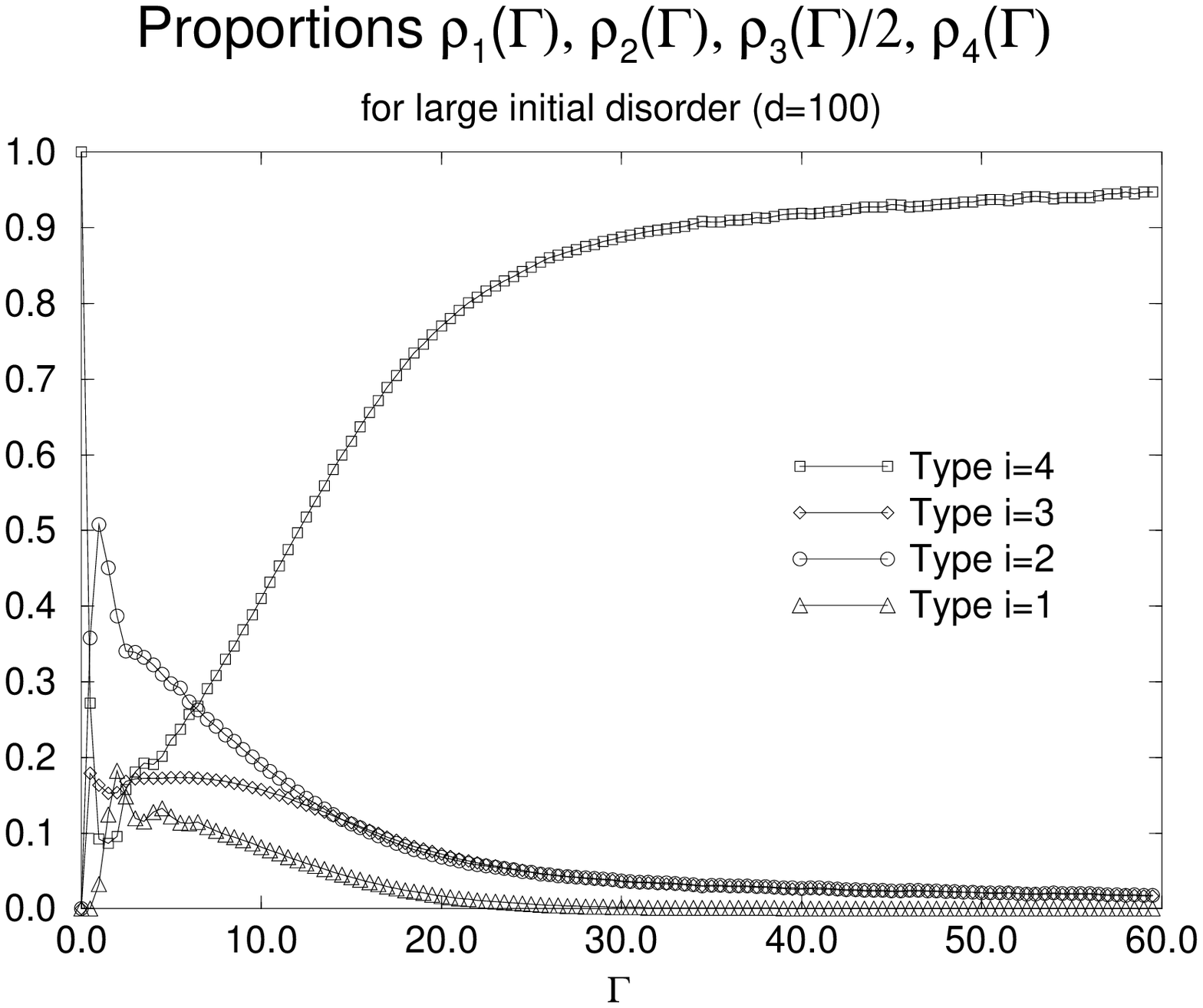}}
\vskip 2.0cm
\caption{
\label{rho100}
{The proportions $\rho_i(\Gamma)$ of the four
types $i=1,2,3,4$ of bonds at scale $\Gamma$, for strong initial
disorder $d=100$ : they reach the asymptotic regime
(\ref{rhofort}).}} 
\end{figure}

\begin{figure}[ht] 
\null
\vglue 5.0cm
\centerline{\epsfbox{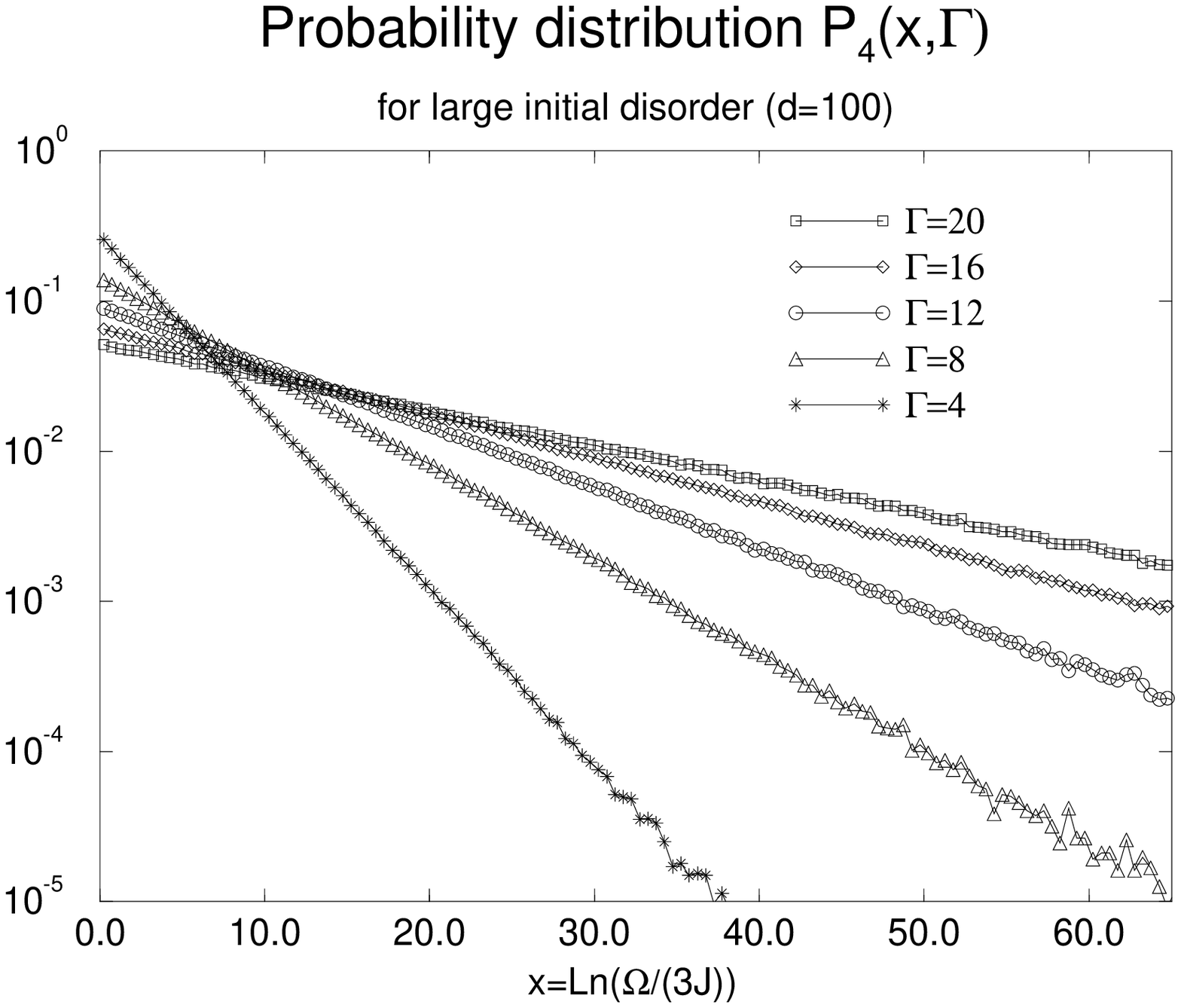}}
\vskip 2.0cm
\caption{
\label{p4d100}
{Linear-Log plot of the probability
distribution ${\cal P}_4 (x, \Gamma)$ for $\Gamma=4,8,12,16,20$, for
strong initial disorder $d=100$ : ${\cal P}_4 (x, \Gamma)$ is well
described by the exponential form (\ref{p4fort}) with a parameter
$\alpha_4(\Gamma)$ plotted on Fig \ref{alpha4d100}.}} 
\end{figure}

\begin{figure}[ht] 
\null
\vglue 5.0cm
\centerline{\epsfbox{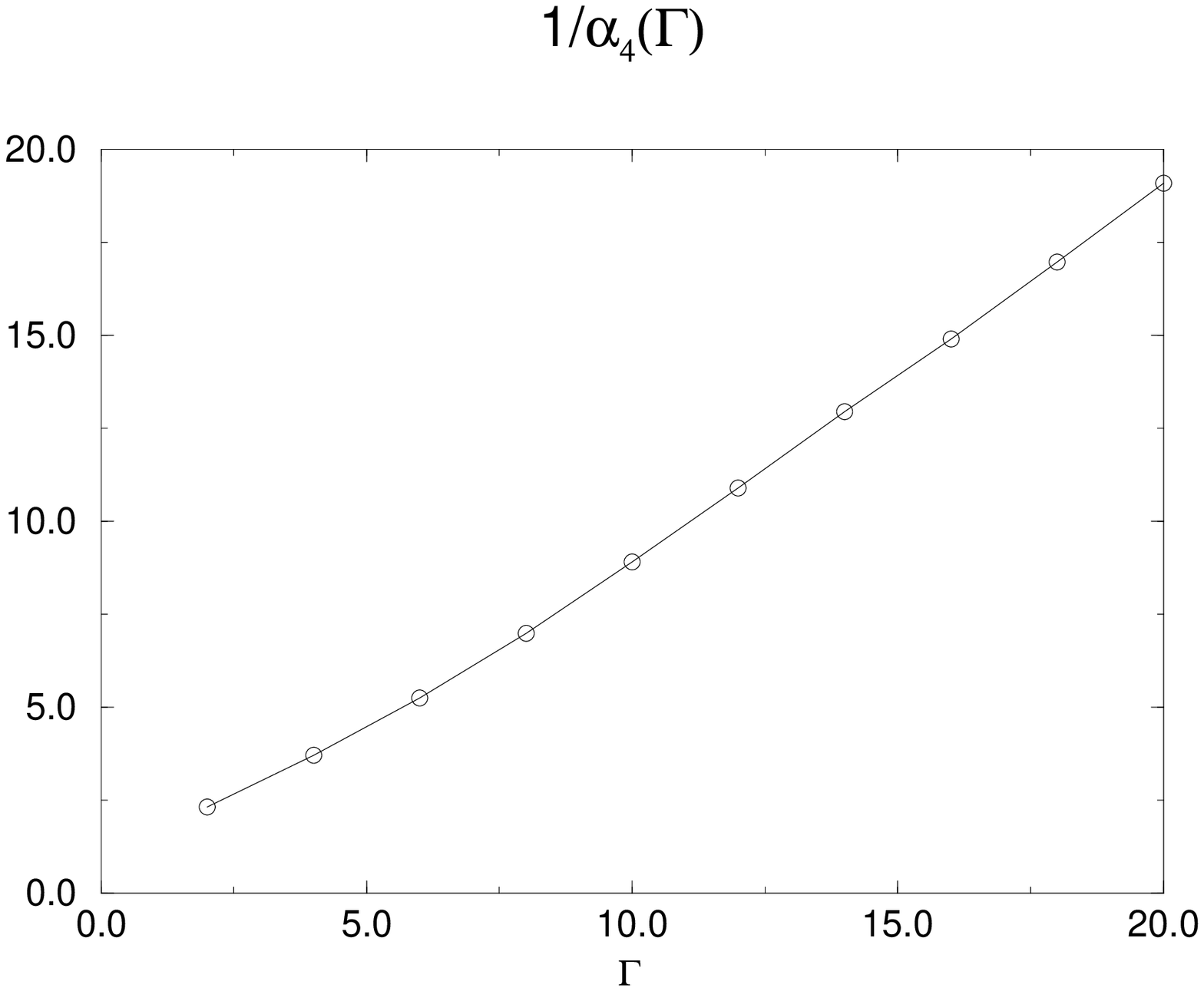}}
\vskip 2.0cm
\caption{
\label{alpha4d100}
{Plot of the inverse of the parameter
$\alpha_4(\Gamma)$ defined in (\ref{p4fort}) as a function of
$\Gamma$ : it follows the random singlet behavior \ref{alphaRS}.}}
\end{figure}

\begin{figure}[ht] 
\null
\vglue 5.0cm
\centerline{\epsfbox{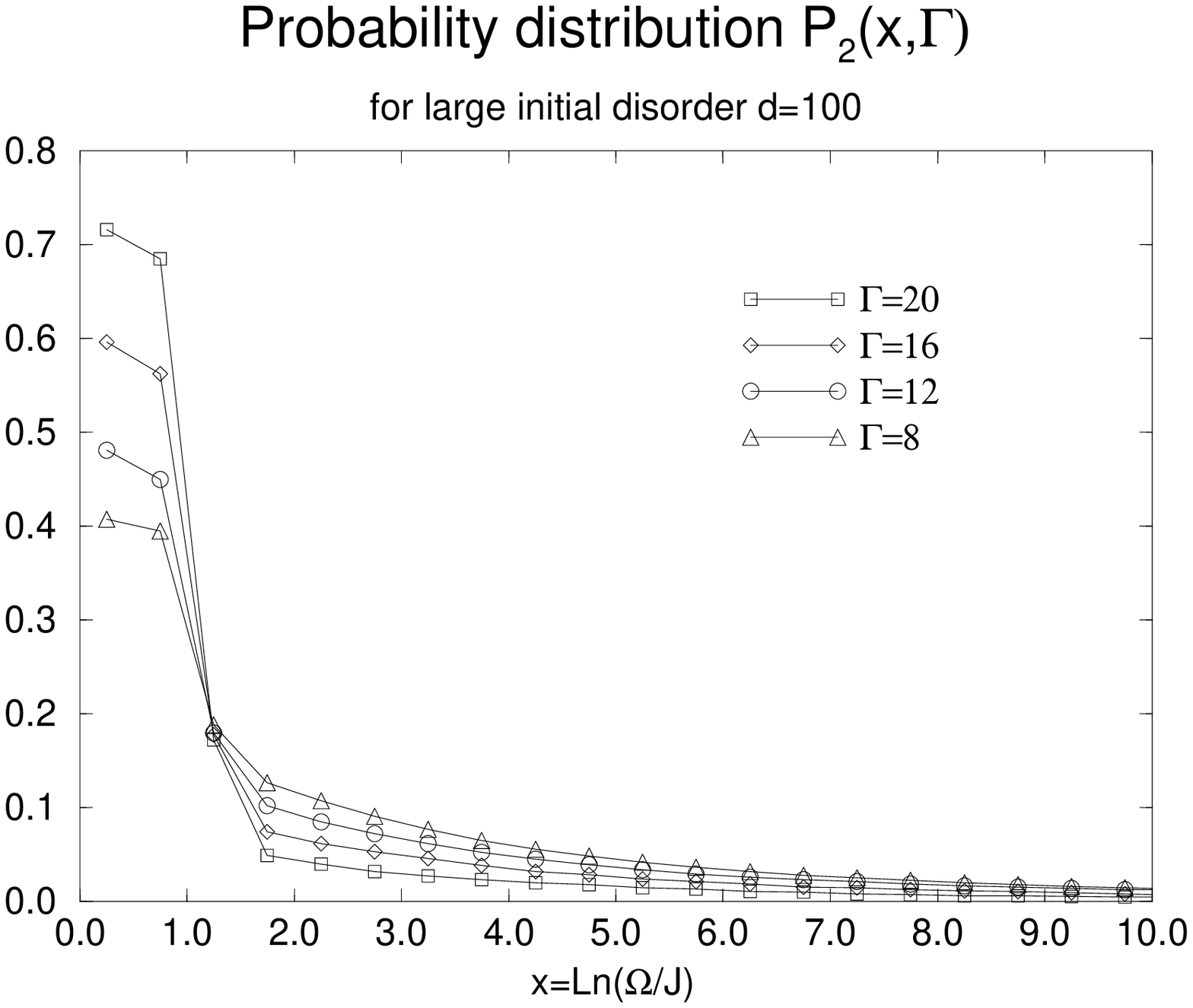}}
\vskip 2.0cm
\caption{
\label{p2d100}
{Plot of the probability distribution ${\cal
P}_2 (x, \Gamma)$ for $\Gamma=8,12,16,20$, for a strong initial
disorder $d=100$ : ${\cal P}_2 (x, \Gamma)$ tends to concentrate on
the interval $0<x< \ln 3$ as explained in the text.}} 
\end{figure}

In the strong disorder phase $d > d_c$, we find that the number
$N(\Gamma)$ of effective spins decays as in the random singlet 
theory
for the disordered antiferromagnetic spin-{1/2} chain (see Fig
\ref{nga100}) :
\be 
N(\Gamma) \oppropto_{\Gamma \to \infty} { 1 \over
\Gamma^2} .
\label{ngafort} 
\ee 
The magnetic susceptibility has thus the random singlet behaviour :
\be
\chi \oppropto {1\over T \log^2 T} .
\ee
The proportions $\rho_i(\Gamma)$ of the
four types of bonds reach an asymptotic regime characterized by (see
Fig \ref{rho100}) 
\be 
\rho_1(\Gamma)\sim 0 \qquad \ \rho_2(\Gamma)
\sim \epsilon(\Gamma) \qquad \ \rho_3(\Gamma) \sim 2 \epsilon(\Gamma)
\qquad \ \rho_4(\Gamma) \sim 1-3\epsilon(\Gamma) 
\label{rhofort} 
\ee
where $\epsilon(\Gamma)$ slowly goes to $0$ as $\Gamma \to \infty$.
This means that there is a sea of bonds of type 4, with sometimes
defects of structure $\{$ bond of type 3,bond of type 2,bond of type
3 $\}$. This defect structure is produced by the renormalization rule
4) for a bond of type 4 when its two neighbor bonds are also of type
4. The fact that there is no more bonds of type 1 in the asymptotic
regime (\ref{rhofort}) shows that defects are destroyed by the
renormalization of the central bond of type 2 and not by the bonds of
type 3 ; this means that for the probability distribution
$P_4(J,\Omega)$ at large enough $\Omega$, two typical couplings are
much weaker than the bigger one. We indeed find that ${\cal P}_4 (x,
\Gamma)$ is an exponential distribution (see Fig \ref{p4d100}) \be
{\cal P}_4 (x, \Gamma) \simeq \alpha_4(\Gamma) e^{- \alpha_4(\Gamma)
x} \ \ \label{p4fort} \ee where the parameter $1/\alpha_4(\Gamma)$
follows the random singlet behavior (see Fig. \ref{alpha4d100}) 
\be
{1 \over \alpha_4(\Gamma) } \simeq \Gamma +\hbox{Cst} 
\label{alphaRS} 
\ee 
As
a consequence, if a defect is produced at the renormalization energy
scale $\Omega$, it survives until the energy scale ${\Omega \over 
3}$
where it get decimated according to the rule (2), and the whole
defect of structure $\{$bond of type 3 , bond of type 2 , bond of
type 3$\}$ entirely disappears to give one bond of type 4. Fig
\ref{p2d100} shows indeed clearly that the probability distribution
${\cal P}_2 (x, \Gamma)$ tends to concentrate on the interval
$0<x<\ln 3$ as $\Gamma$ increases. That has to be contrasted with 
the
bonds of type 3, which are characterized by a distribution ${\cal
P}_3 (x, \Gamma)$ that tends to coincide with ${\cal P}_4 (x,
\Gamma)$ for large enough $\Gamma$.

\subsection{ Approximate flow equations phase}

Assuming that there is a sea of bonds of type 4, with sometimes
defects of structure $\{$ bond of type 3,bond of type 2,bond of type
3 $\}$, it is possible to write approximate flow equations for the
probability distributions of the couplings normalized according to~:
\be 
1= \int_0^{\Omega} dJ \ P_2(J,\Omega) =\int_0^{{2\Omega \over 3}}
dJ \ P_3 (J,\Omega) = \int_0^{{\Omega \over 3}} dJ \ P_4 (J,\Omega) 
.
\ee 
Assuming that the only two important renormalization processes
are the production of the defect structure {bond of type 3 , bond of
type 2 , bond of type 3} by the renormalization rule 4) for a bond of
type 4 when its two neighbor bonds are also of type 4

\bigskip \bigskip \hbox to 450pt {\hfill \vbox{\hsize=10pt
\centerline{$s_0=1$}\par \centerline{$\bullet$}} \kern -7pt \raise
2pt \vtop{\hsize=45pt \centerline{\hrulefill} \par
\centerline{$J_0$}} \kern -7pt \vbox{\hsize=10pt
\centerline{$s_1=1$}\par \centerline{$\bullet$}} \kern -7pt \raise
2pt \vtop{\hsize=45pt \centerline{\hrulefill} \par
\centerline{$J_1={\Omega \over 3}$}} \kern -7pt \vbox{\hsize=10pt
\centerline{$s_2=1$}\par \centerline{$\bullet$}} \kern -7pt \raise
2pt \vtop{\hsize=45pt \centerline{\hrulefill} \par
\centerline{$J_2$}} \kern -7pt \vbox{\hsize=10pt
\centerline{$s_3=1$}\par \centerline{$\bullet$}}
\hfill$\longrightarrow$\hfill \vbox{\hsize=10pt
\centerline{$s_0=1$}\par \centerline{$\bullet$}} \kern -7pt \raise
2pt \vtop{\hsize=45pt \centerline{\hrulefill} \par
\centerline{$J_0$}} \kern -7pt \vbox{\hsize=10pt
\centerline{$s^\prime_1={1\over 2}$}\par \centerline{$\bullet$}}
\kern -7pt \raise 2pt \vtop{\hsize=45pt \centerline{\hrulefill} \par
\centerline{$J_1={\Omega \over 3}$}} \kern -7pt \vbox{\hsize=10pt
\centerline{$s^\prime_2={1\over 2}$}\par \centerline{$\bullet$}}
\kern -7pt \raise 2pt \vtop{\hsize=45pt \centerline{\hrulefill} \par
\centerline{$J_2$}} \kern -7pt \vbox{\hsize=10pt
\centerline{$s_3=1$}\par \centerline{$\bullet$}} \hfill} \bigskip
\bigskip

and the suppression of the defect structure by the decimation rule 2)

\bigskip \bigskip \hbox to 450pt {\hfill \vbox{\hsize=10pt
\centerline{$s_0=1$}\par \centerline{$\bullet$}} \kern -7pt \raise
2pt \vtop{\hsize=35pt \centerline{\hrulefill} \par
\centerline{$J_0$}} \kern -7pt \vbox{\hsize=10pt
\centerline{$s_1={1\over 2}$}\par \centerline{$\bullet$}} \kern -7pt
\raise 2pt \vtop{\hsize=35pt \centerline{\hrulefill} \par
\centerline{$J_1=\Omega$}} \kern -7pt \vbox{\hsize=10pt
\centerline{$s_2={1\over 2}$}\par \centerline{$\bullet$}} \kern -7pt
\raise 2pt \vtop{\hsize=35pt \centerline{\hrulefill} \par
\centerline{$J_2$}} \kern -7pt \vbox{\hsize=10pt
\centerline{$s_3=1$}\par \centerline{$\bullet$}}
\hfill$\longrightarrow$\hfill \vbox{\hsize=10pt
\centerline{$s_0=1$}\par \centerline{$\bullet$}} \kern -7pt \raise
2pt \vtop{\hsize=107pt \centerline{\hrulefill} \par
\centerline{$J^\prime_0={J_0\,J_2\over 2\, \Omega}$}} \kern -7pt
\vbox{\hsize=10pt \centerline{$s_3=1$}\par \centerline{$\bullet$}}
\hfill} \bigskip \bigskip

we obtain the following approximate flow equations for the three
probability distributions~:

\be 
-{{ \partial P_2 (J,\Omega) } \over {\partial \Omega }} =P_2
(\Omega,\Omega) \ P_2 (J,\Omega) + {1 \over 3} P_4 \left({\Omega
\over 3} ,\Omega \right) {N_4(\Omega) \over N_2(\Omega)} \ \bigg[
\delta \left( J-{\Omega \over 3} \right) - P_2 (J,\Omega) \bigg]
\label{flow2} \ee \be -{{ \partial P_3 (J,\Omega) } \over {\partial
\Omega }} = {2 \over 3} P_4 \left( {\Omega \over 3} ,\Omega \right)
{N_4(\Omega) \over N_3(\Omega)} \ \bigg[ P_4 (J,\Omega)- P_3
(J,\Omega) \bigg] \label{flow3} \ee \bea -{{ \partial P_4 (J,\Omega)
} \over {\partial \Omega }} ={1 \over 3} P_4 \left({\Omega \over 3}
,\Omega \right) \ P_4 (J,\Omega) -{N_2(\Omega) \over N_4(\Omega)} \
P_2 (\Omega,\Omega) \ P_4 (J,\Omega) \\ + {N_2(\Omega) \over
N_4(\Omega)} P_2 (\Omega ,\Omega) \int_0^{{2\Omega \over 3}} dJ_0 \
P_3 (J_0,\Omega) \int_0^{{2\Omega \over 3}} dJ_2 \ P_3 (J_2,\Omega) \
\delta \left( J-{{J_0 J_2} \over {2\, \Omega}}\right) ,
\label{flow4}
\eea 
together with the flow equations for the number $N_i(\Omega)$ of
bonds of type $i=2,3,4$~:
\be 
-{{dN_2} \over {d\Omega}} = - { 1 \over
2} {{dN_3} \over {d\Omega}} ={1 \over 3} P_4 \left({\Omega \over 3}
,\Omega \right)\ N_4 (\Omega) - P_2(\Omega,\Omega) \ N_2(\Omega) \ee
\be -{{dN_4} \over {d\Omega}} = P_2(\Omega,\Omega) \ N_2(\Omega) -
P_4 \left({\Omega \over 3} ,\Omega \right) \ N_4 (\Omega) ,
\ee 
so that
the total number $N(\Omega)=N_2(\Omega)+N_3(\Omega)+N_4(\Omega)$ of
bonds evolves according to~:
\be 
-{{dN} \over {d\Omega}} = - 2
P_2(\Omega,\Omega) \ N_2(\Omega) .
\label{flotN} 
\ee

It is more convenient to write the flow equations for the probability
distributions ${\cal P}_i (x, \Gamma)$ of the reduced variable $x=\ln
\left({\Omega \over{\Delta(J)}} \right)$, where $\Delta(J)$ is
defined by (\ref{defxtype}) so that the random variable $x$ varies in
$(0,\infty)$ for any type of bonds~:
\be 
{{ \partial {\cal
P}_2(x,\Gamma) } \over {\partial \Gamma }} ={{ \partial {\cal
P}_2(x,\Gamma) } \over {\partial x }} +{\cal P}_2 (0,\Gamma) \ {\cal
P}_2 (x,\Gamma) \\ + {N_4(\Gamma) \over N_2(\Gamma)} \ {\cal P}_4
(0,\Gamma) \ \bigg[ \delta \left( x- \ln 3 \right) - {\cal P}_2
(x,\Gamma)\bigg] ,
\label{flotp2} 
\ee 
\be 
{{ \partial {\cal
P}_3(x,\Gamma) } \over {\partial \Gamma }} ={{ \partial {\cal
P}_3(x,\Gamma) } \over {\partial x }} + 2 {N_4(\Gamma) \over
N_2(\Gamma)} \ {\cal P}_4 (0,\Gamma) \ \bigg[ {\cal P}_4 (x- \ln
2,\Gamma) - {\cal P}_3 (x,\Gamma)\bigg] ,
\ee 
\bea 
{{ \partial {\cal
P}_4(x,\Gamma) } \over {\partial \Gamma }} ={{ \partial {\cal
P}_4(x,\Gamma) } \over {\partial x }} + \bigg[ {\cal P}_4 (0,\Gamma)
\ - {N_2(\Gamma) \over N_4 (\Gamma)} \ {\cal P}_2 (0,\Gamma) \bigg]
{\cal P}_4 (x,\Gamma) \\ + {N_2(\Gamma) \over N_4(\Gamma)} {\cal P}_2
(0,\Gamma) \int_0^{\infty} dx_1 \ {\cal P}_3 (x_1,\Gamma)
\int_0^{\infty} dx_2 \ {\cal P}_3 (x_2,\Gamma) \ \delta \left(
x-x_1-x_2 -\ln {3 \over 2} \right) .
\eea

Since the singular term containing the delta-function in
(\ref{flotp2}) tends to develop a discontinuity in ${\cal
P}_2(x,\Gamma)$ at $x=\ln 3$, it is convenient to set~:
\be 
{\cal
P}_2(x,\Gamma) =\big[ 1- e(\Gamma)\big] { {\theta( \ln 3 -x )} \over
{\ln 3}} + e(\Gamma) f_2(x,\Gamma) ,
\ee 
where $0<e(\Gamma)<1$ and
$f_2(x,\Gamma)$ is a normalized probability distribution that is
regular at $x=\ln 3$. Equation (\ref{flotp2}) will be satisfied if
$e(\Gamma)$ and $f(x,\Gamma)$ satisfy~:
\bea 
e(\Gamma)=1- ( \ln 3 ) \
{N_4(\Gamma) \over N_2(\Gamma)} \ {\cal P}_4 (0,\Gamma) \\ {
{de(\Gamma)} \over {d \Gamma}} = - e(\Gamma) \big[1- e(\Gamma) ]
f(0,\Gamma) \\ {{ \partial f(x,\Gamma) } \over {\partial \Gamma }}
={{ \partial f(x,\Gamma) } \over {\partial x }} +f (0,\Gamma) \ f
(x,\Gamma) .
\eea 
Obvious stationary solutions for $f(x,\Gamma)$ are
simple exponentials~:
\be 
f(x,\Gamma) \opsimeq_{\Gamma \to \infty}
\alpha_f \ e^{- \alpha_f x} ,
\ee 
in which case $e(\Gamma)$ vanishes
exponentially~:
\be 
e(\Gamma) \oppropto_{\Gamma \to \infty} e^{-
\alpha_f \Gamma} ,
\ee 
so that ${\cal P}_2(x,\Gamma)$ converges towards
the stationary solution~:
\be 
{\cal P}_2^*(x) = { 1 \over {\ln 3}} \
\theta( \ln 3 -x ) .
\ee 
This corresponds in the original variables to
\be 
P_2 (J,\Omega) = { 1 \over {(\ln 3 ) J }} 
\qquad \hbox{for}
{\Omega \over 3} < J < \Omega .
\ee 
We also obtain the following
equation in the asymptotic regime~:
\be 
N_4 (\Gamma) \ {\cal P}_4
(0,\Gamma) \simeq N_2(\Gamma) \ { 1 \over {\ln 3}} ,
\label{balance}
\ee 
that we will use now to study the flow equations for ${\cal
P}_3(x,\Gamma)$ and ${\cal P}_4(x,\Gamma)$

With the last change of variables~:
\be 
x \longrightarrow z=\alpha_4
(\Gamma) x ,
\ee 
we find that the flow equation for the corresponding
probability distributions $\tilde P_4(z, \Gamma)$ and $\tilde P_3(z,
\Gamma)$ admit the stationary solutions~:
\be 
\tilde P_4(z, \Gamma)
\operarrow_{ \Gamma \to \infty} e^{-z} \qquad \hbox{and} \qquad
\tilde P_3(z, \Gamma) \operarrow_{ \Gamma \to \infty} e^{-z} ,
\ee
where 
\be 
\alpha_4(\Gamma) \oppropto_{\Gamma \to \infty} { 1 \over
\Gamma } ,
\ee 
as in the random singlet solution of Ma-Dasgupta. It is
then easy to obtain the asymptotic behavior of the total number
$N(\Gamma)$ of spins (\ref{flotN})~:
\be 
N(\Gamma) \oppropto_{\Gamma
\to \infty} { 1 \over \Gamma^2 } ,
\ee 
and the asymptotic behavior of
the proportion $\epsilon(\Gamma)$ (\ref{rhofort}) of defects from
(\ref{balance})~:
\be 
\epsilon(\Gamma) ={N_2(\Gamma) \over N (\Gamma)}
\oppropto_{\Gamma \to \infty} { {\ln 3} \over \Gamma } .
\ee

\section{The  critical regime}

\begin{figure}[ht] 
\null
\vglue 5.0cm
\centerline{\epsfbox{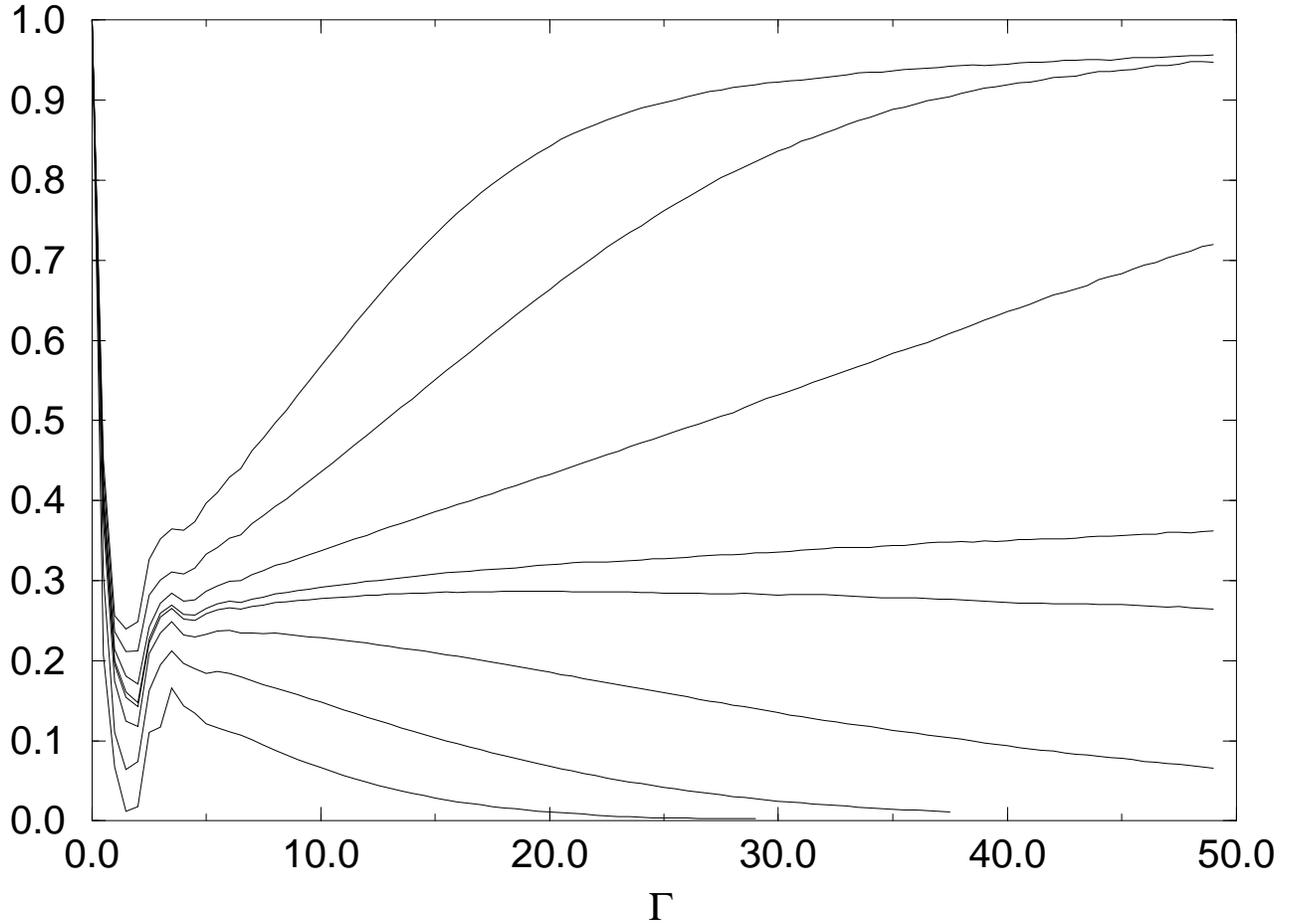}}
\vskip 2.0cm
\caption{
\label{s1global}
{Proportion of spins $S=1$ among the
effective spins at scale $\Gamma$ for various values of the initial
disorder $d=1,2,3,4,5.5,6,8,16,100$ : this proportion flows towards
$0$ in the weak disorder phase and to $1$ in the strong disorder
phase. Between these two attractive values, there is an unstable
fixed point at $d_c \simeq 5.75(5)$ where the proportion of spins
$S=1$ among the effective spins remains stationary at the
intermediate value $0.315(5)$.}} 
\end{figure}

\begin{figure}[ht] 
\null
\vglue 3.0cm
\centerline{\epsfbox{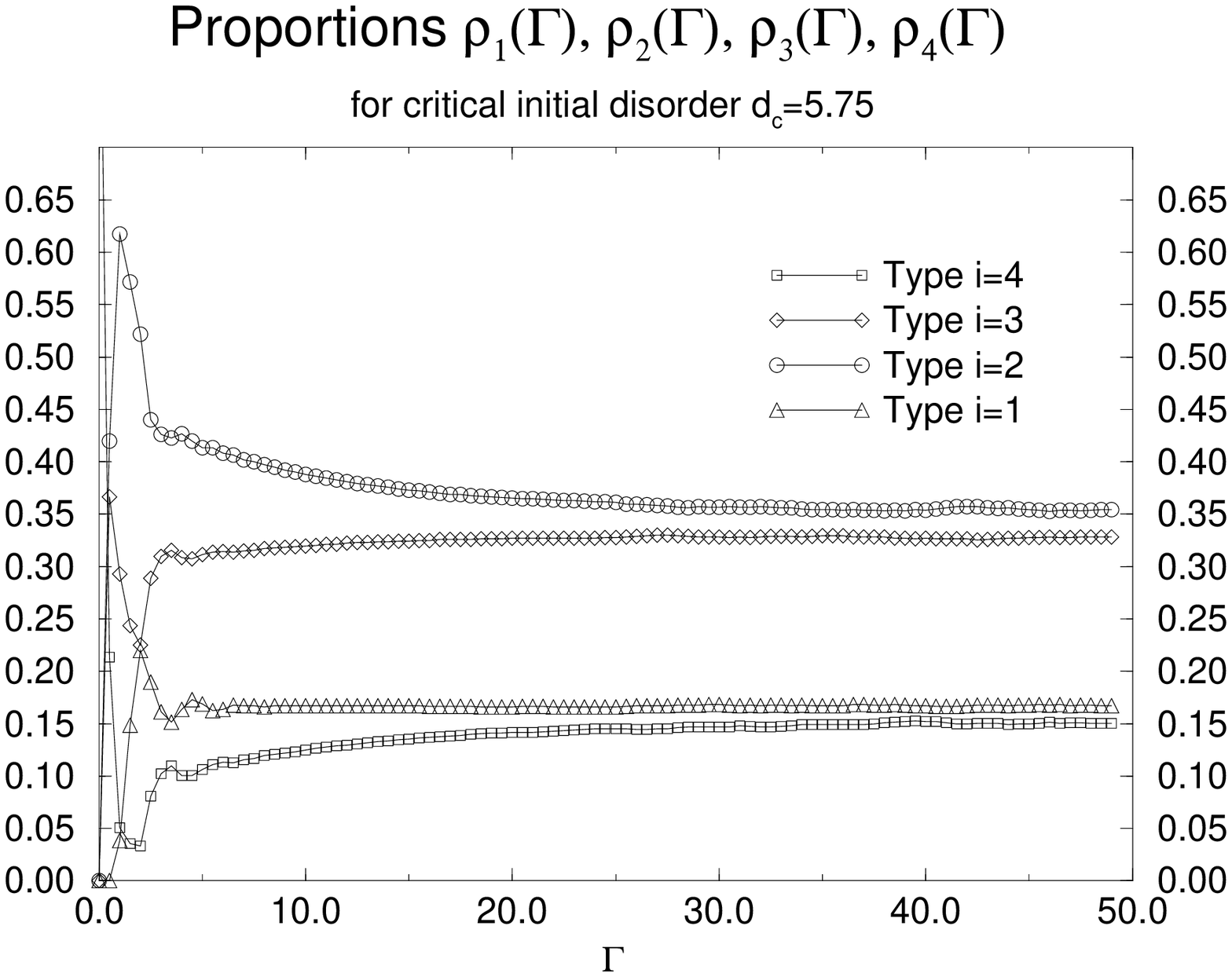}}
\vskip 1.5cm
\caption{
\label{rhocrit}
{ The proportions $\rho_i(\Gamma)$ of the
four types $i=1,2,3,4$ of bonds at scale $\Gamma$, for the critical
initial disorder $d_c=5.75$ (\ref{rhocnume}).}} 
\end{figure}

\begin{figure}[ht] 
\null
\vglue 5.0cm
\centerline{\epsfbox{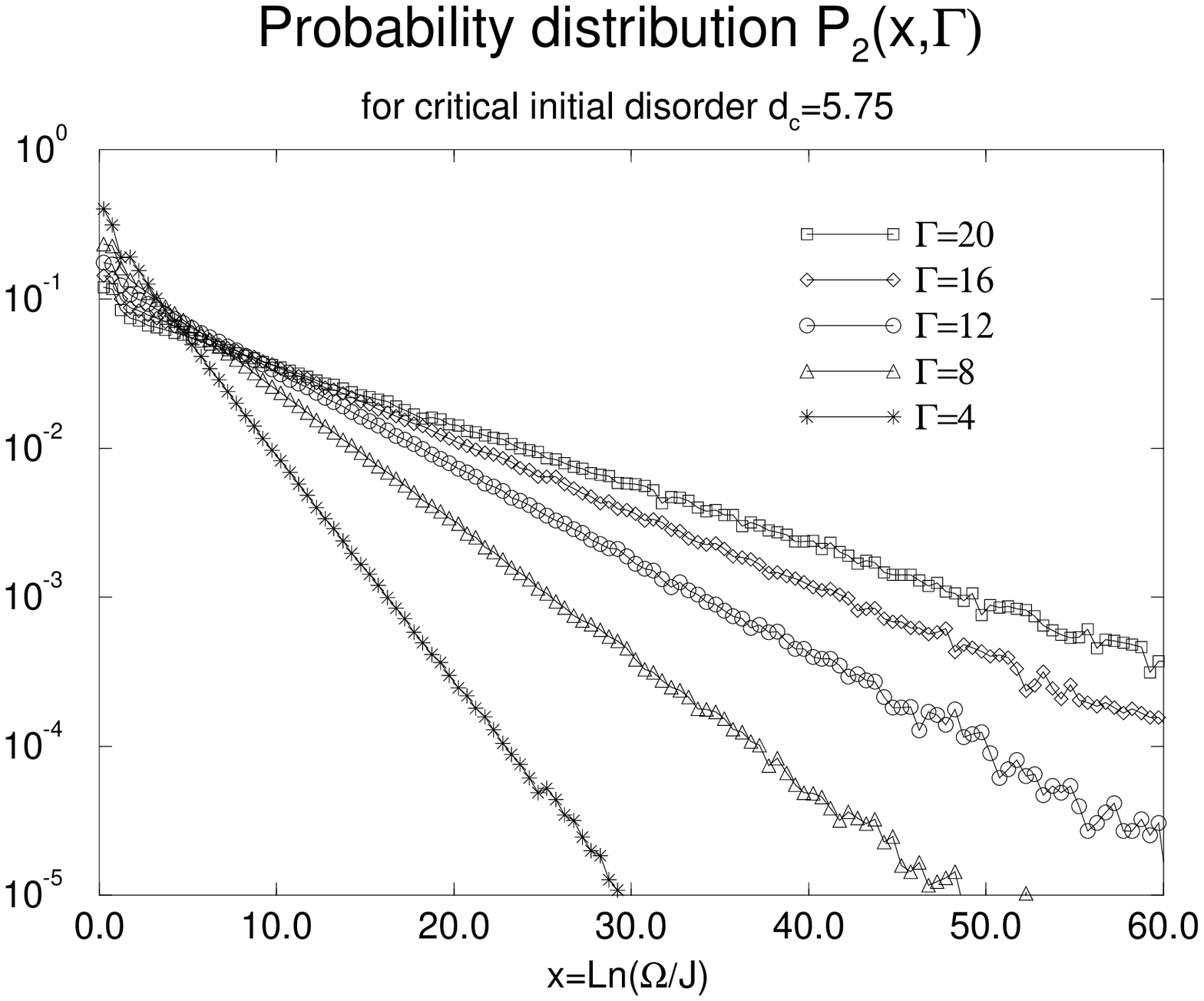}}
\vskip 2.0cm
\caption{
\label{p2crit}
{Linear-Log plot of the probability
distribution ${\cal P}_2 (x, \Gamma)$ for $\Gamma=4,8,12,16,20$, for
critical initial disorder $d_c=5.75$ : ${\cal P}_2 (x, \Gamma)$ takes
the exponential form (\ref{pcritnume}) with a parameter
$\alpha_c(\Gamma)$ plotted on Fig \ref{alphac}.}} 
\end{figure}

\begin{figure}[ht] 
\null
\vglue 5.0cm
\centerline{\epsfbox{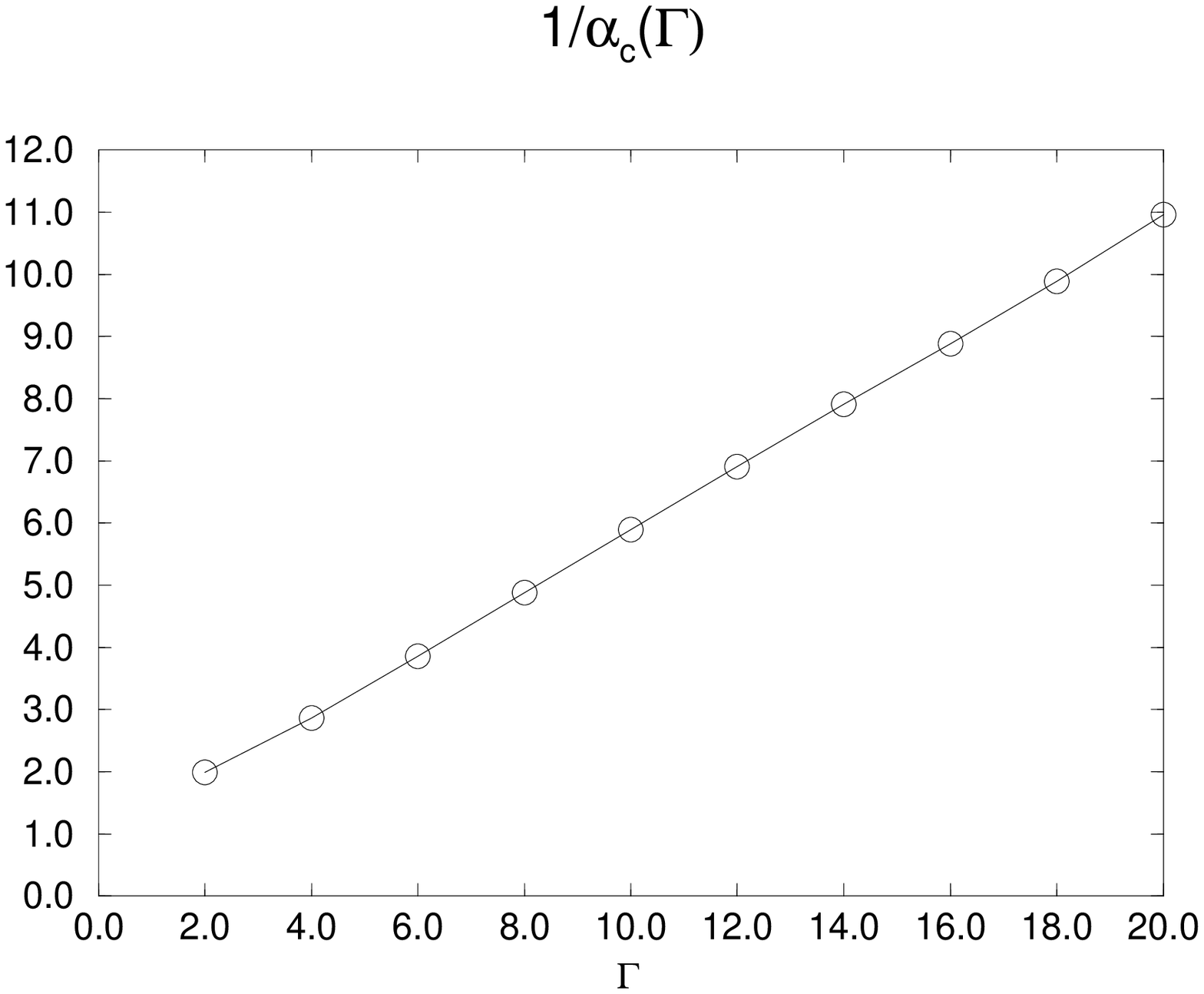}}
\vskip 2.0cm
\caption{
\label{alphac}
{Plot of the inverse of the parameter
$\alpha_c(\Gamma)$ defined in (\ref{pcritnume}) as a function of
$\Gamma$ : it follows the behavior (\ref{alphacnume}).}} \end{figure}

On Fig. \ref{s1global}, we have plotted the proportion $ { {
N_{(S=1)} (\Gamma)} \over { N (\Gamma)} } $ of spins $S=1$ among the
effective spins for various values of the disorder this proportion
flows towards $0$ in the weak disorder phase and to $1$ in the strong
disorder phase. Between these two attractive values, there is an
unstable fixed point at $d_c \simeq 5.75(5)$ where the proportion of
spins $S=1$ among the effective spins remains stationary at the
intermediate value $0.315(5)$. The proportions $\rho_i(\Gamma)$ of
the four types of bonds reach a stationary state characterized by
(see Fig \ref{rhocrit}) 
\be 
\rho_1(\Gamma)\sim 0.17 \, ,\qquad \
\rho_2(\Gamma) \sim 0.35 \, ,\qquad \ 
\rho_3(\Gamma) \sim 0.33 \, ,\qquad \
\rho_4(\Gamma) \sim 0.15 \, .
\label{rhocnume} 
\ee 
We find of course that the four probability distributions ${\cal
P}_i(x,\Gamma)$ for $i=1,2,3,4$ coincide up to statistical
fluctuations (otherwise, the proportions $\rho_i(\Gamma)$ would not
remain stationary) and follow the exponential form (see Fig
\ref{p2crit})~:
\be 
{\cal P}_i (x, \Gamma) \simeq \alpha_c(\Gamma) e^{-
\alpha_c(\Gamma) x} \ \ \label{pcritnume} ,
\ee 
where the parameter
$1/\alpha_c(\Gamma)$ (see Fig. \ref{alphac}) follows the behavior of
the effective model of Hyman and Yang \cite{HY} :
\be 
{1 \over
\alpha_c(\Gamma) } \simeq {\Gamma \over 2} +\hbox{Cst} . 
\label{alphacnume}
\ee
The magnetic susceptibility is given by the effective number of free 
spins~:
\be
\chi \oppropto {1\over T \log^3 T}.
\ee

\section{Conclusion}

We have introduced a real-space renormalization scheme
that allows the study of the spin-1 chain. 
Within this scheme we obtained a complete characterization
of the weak-coupling phase, the critical regime and
the strong-disorder phase. In all phases we were able
to follow the spin populations and to obtain the
probability distributions of the different types of bonds
that appear under renormalization. It is only in the weak
and strong coupling limit that we were able to obtain
approximate analytical flow equations.

The renormalization scheme that we used is an
extension of the Ma-Dasgupta idea. These schemes
have in common the fact that they are consistent 
for arbitrarily weak 
initial disorder. They do not create bonds stronger than
the original decimated bond.
In the spin-1/2 case, it is believed that
this means that there is no critical disorder. In fact,
this is suggested by bosonization: most bosonic forms 
of randomness give rise to relevant operators
along the massless line of the pure system when the anisotropy
is varied. The simplest assumption\cite{Fis94} is thus that
the system flows immediately to the random-singlet phase
(there is a region of stability of the spin liquid
but this happens only for attractive enough interactions
between the Jordan-Wigner fermions i.e. for negative
enough anisotropy).

However, this is not the case for the spin-1 chain. Here
the Haldane gap is perturbatively insensitive
to disorder as naively expected. This is known from bosonization
studies of the spin-1/2 two-leg ladder\cite{Ori} as well
as of the anisotropic spin-1 chain\cite{VB}. So we may be in a 
situation with a first critical disorder strength
corresponding to the vanishing of Haldane gap
but which is unreachable by the real-space scheme. 
With increasing disorder there is then the second
critical disorder strength for which the string order
vanishes. This second transition is described by 
our renormalization scheme which is then asymptotically
exact. Conversely,
the bosonization methods are unable to follow the flow
to strong coupling and thus are unable to describe even
the weak-disorder phase captured by the real-space scheme.
It may be also that there is nothing like a critical value
of the disorder for the vanishing of the Haldane gap,
if for example there are states of arbitrarily small
energies in the gap as in the case of the Lifshitz tails
in the localization problem.
It remains to be seen if there is a single theoretical approach
that is able to deal all known limiting cases.



\end{document}